\documentclass[conference]{IEEEtran}
\usepackage[comma,numbers,square,sort&compress]{natbib}
\usepackage{amsmath}
\usepackage{amssymb}
\usepackage{float}
\usepackage{amsthm}
\usepackage{graphicx}
\usepackage{subfigure}
\usepackage{epstopdf}
\usepackage{color}
\usepackage{verbatim}
\usepackage{tikz,times}
\usepackage{algorithmic,algorithm}

\newtheorem{defn}{Definition}

\newtheorem{thm}{Theorem}
\newtheorem{lem}{Lemma}
\newtheorem{assum}{Assumption}

\ifCLASSINFOpdf
\else
\fi
\begin{document}
%
\title{Distributed Kalman  Filter for A Class of Nonlinear Uncertain Systems: An Extended State  Method}

\author{\IEEEauthorblockN{Xingkang He, Xiaocheng Zhang, Wenchao Xue*, Haitao Fang}
\IEEEauthorblockA{LSC, NCMIS, Academy of Mathematics and Systems Science,\\
	Chinese Academy of Sciences, Beijing 100190, China. \\
 School of Mathematical Sciences, University of Chinese Academy of Sciences, Beijing 100049, China.\\
 Email: xkhe@amss.ac.cn, zhangxiaocheng16@mails.ucas.ac.cn, wenchaoxue@amss.ac.cn*, htfang@iss.ac.cn}
\and


}


%


\maketitle

\begin{abstract}
This paper studies the distributed state estimation problem for a class of discrete-time stochastic systems with nonlinear uncertain dynamics over time-varying topologies of sensor networks.
An extended state vector consisting of the original state and the nonlinear dynamics  is constructed.  By analyzing the extended system, we
provide a  design method for the filtering gain and fusion matrices, leading to the extended state  distributed Kalman filter. It is shown that the proposed filter can provide the upper bound of estimation covariance in real time, which means the estimation accuracy can be evaluated online. 
It is proven that the  estimation covariance of the filter is bounded under rather mild assumptions, i.e., collective observability of the system and jointly strong connectedness of network topologies.
 Numerical simulation shows the effectiveness of the proposed filter.
\end{abstract}


%
\IEEEpeerreviewmaketitle

\section{Introduction}
%
%

In recent years, networked state estimation problems are drawing more and more attention of researchers, due to the broad applications in environmental monitoring, target tracking over networks, collaborative information processing, etc.  Two main strategies, namely, the centralized and the distributed, have been considered   to deal with the state estimation problems over networks. The Kalman filter based fusion estimation problems were studied in  \cite{Pfaff2017Information,Chong2014Comparison}, where each sensor can fuse the information from all the other sensors over the network to obtain the better estimate. In the distributed strategy, the communications between sensors follow a peer-to-peer protocol. 
Compared with the centralized strategy, the distributed one  has shown more robustness in network structure, better performance in energy saving and stronger ability in parallel processing. 

In the existing literature on distributed state estimation, many effective methods and some theoretical analysis tools have been provided.
A  distributed estimator with time-invariant filtering gain was studied in \cite{Khan2014Collaborative}, where the  
relationship between the instability of system and the boundedness of estimation error was evaluated.
A distributed Kalman filter (DKF) based on measurement consensus strategy was proposed in \cite{Das2015Distributed}, where the design method of the consensus weights and the filtering gain were investigated. In \cite{Cat2010Diffusion}, a diffusion DKF with constant weights was studied and the performance of the proposed filter was analyzed theoretically. Yet, the results were based on a local observability condition \footnote{ Each sub-system on an individual sensor is observable.},   which is difficult to be met in a large network. Besides, there were many methods and results related to the distributed state estimation. Interested readers can refer to 
\cite{Mahmoud2013Distributed,He2017Consistent,Battistelli2015Consensus,yang2017stochastic} and the references therein.

The state estimation problems for nonlinear systems have been studied for a few decades. 
Through linearizing the nonlinear system to the first order, the extended Kalman filter (EKF) was constructed in the form of Kalman filter \cite{Reif1999Stochastic} to achieve the state estimation and prediction. Based on the third-order linearization for any known nonlinearity, unscented Kalman filter (UKF) can accurately achieve the estimation on the posterior mean and covariance \cite{Julier2004Unscented}.   
For the nonlinear uncertain systems, quite a few robust filters including $H_{\infty}$ filters and set valued filters have been studied by many researchers \cite{yang2008non,ding2012distributed,calafiore2005reliable}. The essential idea was to minimize a worst-case bound so as to obtain a conservative estimate of the  state under deterministic or stochastic uncertainties of the system. Due to the instability issue of the linearization methods and the conservativeness of the robust filters, \cite{Bai2017reliable} proposed an extended state Kalman filter to handle the state estimation for a class of nonlinear time-varying uncertain systems.  Moreover, the upper boundedness of the estimation covariance and the  optimality of estimation in some certain conditions are discussed.
The corresponding continuous-time problem was studied in \cite{Zhang2018}, which proposed an extended state Kalman-Bucy filter and analyzed the main performance of the filter.
 While, all the above filters still face many problems when they are utilized to the sensor networks, especially in the aspects of stability analysis and performance evaluation. Thus, more attention should be paid to the research of distributed state estimation for nonlinear uncertain systems.


In this paper, we study the distributed state estimation for a class of discrete-time stochastic systems with nonlinear uncertain dynamics  over time-varying topologies of sensor networks. The main contributions of this paper are summarized. First, a new extended state method is provided and the method can achieve the simultaneous estimation of the original state and the nonlinear dynamics. By designing adaptive fusion matrices and filtering gain, the upper bounds of estimation covariances  can be obtained in real time. Second, it is proven that the  estimation covariance of the filter is bounded under rather mild assumptions, i.e., collective observability of system and jointly strong connectedness of network topologies.

The remainder of the paper is organized as follows: Section \uppercase\expandafter{\romannumeral2} is on the preliminaries and problem formulation. Section \uppercase\expandafter{\romannumeral3} presents the main results of this paper.
Section \uppercase\expandafter{\romannumeral4} 
shows the numerical simulation. The conclusion of this paper is given in Section \uppercase\expandafter{\romannumeral5}. Due to the limitation of the pages, the proofs are omitted. Interested readers can refer to the full version in \cite{He2018Dis} for the detailed proofs.

\subsection{Notations}
The superscript ``T" represents the transpose. $I_{n}$ stands for the identity matrix with $n$ rows and $n$ columns. $E\{x\}$ denotes the mathematical expectation of the stochastic variable $x$, and  $blockdiag\{\cdot\}$ means the block elements are arranged in diagonals. $diag\{\cdot\}$ represent the diagonalization scalar elements. $tr(P)$ is the trace of the matrix $P$. $\mathbb{N}^+$ denotes the set of positive natural numbers. $\mathbb{R}^n$ stands for the set of $n$-dimensional real vectors.

\section{Preliminaries and Problem Formulation}
\subsection{Preliminaries in Graph Theory}
Let $\mathcal{G}=(\mathcal{V},\mathcal{E}_\mathcal{G},\mathcal{A}_\mathcal{G})$ be a weighted digraph, which consists of  the node set  $\mathcal{V}=\{1,2,\cdots,N\}$, the set of edges $\mathcal{E}_\mathcal{G}\subseteq \mathcal{V}\times \mathcal{V}$ and the weighted adjacent matrix $\mathcal{A}_\mathcal{G}=[a_{i,j}]\in \mathbb{R}^{N\times N}$. In the weighted adjacent matrix $\mathcal{A}_\mathcal{G}$, all the elements are nonnegative, row stochastic and the diagonal elements are all positive, i.e., $a_{i,i}> 0,a_{i,j}\geq 0,\sum_{j\in \mathcal{V}}a_{i,j}=1$. If $a_{i,j}>0,j\neq i$, there is a link $(i,j)\in \mathcal{E}_\mathcal{G}$, which means node $i$ can directly receive the information of node $j$ through the communication channel. In this situation, node $j$ is called the neighbor of node $i$ and all the neighbors of node $i$ including itself can be represented by the set $\{j\in\mathcal{V}|(i,j)\in \mathcal{E}_\mathcal{G}\}\bigcup\{i\}\triangleq \mathcal{N}_{i}^\mathcal{G}$, whose size  is denoted as $|\mathcal{N}_{i}^\mathcal{G}|$.
For a given positive integer $L$, the union of the $L$ digraphs $\mathcal{G}_1=(\mathcal{V},\mathcal{E}_{\mathcal{G}_1},\mathcal{A}_{\mathcal{G}_1}),\cdots,\mathcal{G}_L=(\mathcal{V},\mathcal{E}_{\mathcal{G}_L},\mathcal{A}_{\mathcal{G}_L})$  is denoted as $\sum_{l=1}^{L}\mathcal{G}_l=(\mathcal{V},\sum_{l=1}^{L}\mathcal{E}_{\mathcal{G}_l},\sum_{l=1}^{L}\mathcal{A}_{\mathcal{G}_l})$.
$\mathcal{G}$ is called strongly connected if for any pair nodes $(i_{1},i_{l})$, there exists a direct path from $i_{1}$ to $i_{l}$ consisting of edges $(i_{1},i_{2}),(i_{2},i_{3}),\cdots,(i_{l-1},i_{l})$. We call $\{\mathcal{G}_1,\cdots,\mathcal{G}_L\}$  jointly strongly connected if $\sum_{l=1}^{L}\mathcal{G}_l$ is strongly connected.

\subsection{Problem Formulation}

Consider the following discrete-time stochastic system
\begin{equation}\label{system}
\begin{cases}
x_{k+1}=\bar A_{k}x_{k}+\bar G_{k}F(x_{k},k)+\bar \omega_{k},\\
y_{k,i}=\bar H_{k,i}x_{k}+v_{k,i},i=1,2,\cdots,N,
\end{cases}
\end{equation}
where $x_{k}$ is the unknown $n$-dimensional system state, $\bar A_{k}$ is the known system matrix and $\bar\omega_{k}$ is the zero-mean unknown process noise with $E\{\bar\omega_{k}\bar\omega_{k}^T\}>0$.
$F(x_{k},k)$ is the  $p$-dimensional uncertain dynamics consisting of the known nominal model $\bar F(x_{k},k)$ and some unknown disturbance. For convenience, we define $F(x_{k},k)\triangleq F_k$ and $\bar F(x_{k},k)\triangleq\bar  F_k$. $N$ is the number of sensors over the network.
$y_{k,i}$ is the $m_i$-dimensional measurement vector obtained via sensor $i$, $\bar H_{k,i}$ is the known measurement matrix,  and $v_{k,i}$ is the zero-mean stochastic measurement noise.  $\{\bar \omega_{k}\}_{k=0}^{\infty}$, $\{v_{k,i}\}_{k=0}^{\infty}$ are independent of each other, and also independent of $x_0$ and $F_0$.
It is noted that  $\bar A_{k}$, $\bar G_{k}$, $\bar H_{k,i}$, and $y_{k,i}$ are simply known to sensor $i$. The above matrices and vectors have compatible dimensions.


For the system (\ref{system}), a new state vector, consisting of the original state $x_{k}$ and the nonlinear uncertain dynamics $F_{k}$, can be constructed. Then a modified system model with respect to the new state vector is given in the following.
\begin{align}\label{system2}
\begin{cases}
\begin{pmatrix}
x_{k+1} \\
F_{k+1}
\end{pmatrix}=\begin{pmatrix}
\bar A_{k} & \bar G_{k}\\
0 & I_p 
\end{pmatrix}\begin{pmatrix}
x_{k} \\
F_{k}
\end{pmatrix}+\begin{pmatrix}
\bar \omega_{k} \\
0 
\end{pmatrix}\\
\qquad\qquad\qquad+\begin{pmatrix}
0 \\
I_p
\end{pmatrix}(F_{k+1}-F_{k}),\\
y_{k,i}=\begin{pmatrix}
\bar H_{k,i} & 0 
\end{pmatrix}\begin{pmatrix}
x_{k} \\
F_{k}
\end{pmatrix}+v_{k,i}.
\end{cases}
\end{align}

Define
\begin{align}\label{new_matrices}
\begin{cases}
\begin{pmatrix}
x_{k} \\
F_{k} 
\end{pmatrix}\triangleq X_{k},
\begin{pmatrix}
\bar A_{k} & \bar G_{k} \\
0 & I_p 
\end{pmatrix}\triangleq A_{k},\\
\begin{pmatrix}
\bar \omega_{k} \\
0 
\end{pmatrix}\triangleq\omega_{k},\begin{pmatrix}
0 \\
I_p
\end{pmatrix}\triangleq D,\\
F_{k+1}-F_{k}\triangleq u_{k},\\
\begin{pmatrix}
\bar H_{k,i} & 0 
\end{pmatrix}\triangleq H_{k,i},
\end{cases}
\end{align}
then the system (\ref{system2}) can be rewritten as
\begin{equation}\label{system3}
\begin{cases}
X_{k+1}= A_{k}X_{k}+Du_{k}+\omega_{k},\\
y_{k,i}= H_{k,i}X_{k}+v_{k,i}.
\end{cases}
\end{equation}

In this paper, the following assumptions are needed.
\begin{assum}\label{ass_sys}
	The system (\ref{system3}) satisfies the following conditions
	\begin{equation}
	\begin{split}
		& E\{\omega_{k}\omega_{k}^T\}\leq Q_{k},\\
	&E\{v_{k,i}v_{k,i}^T\}\leq R_{k,i},i=1,2,\cdots,N,\\\
	& E\{u_{k}^2(j)\}\leq q_{k}(j), j=1,2,\cdots,p,
	\end{split}
	\end{equation}
	where $R_{k,i}>0$ and $q_{k}(j)>0$. Also, $Q_{k}$ and $q_{k}(j)$ are uniformly upper bounded.
\end{assum}

The  noise conditions given in Assumption \ref{ass_sys} are reasonable and easy to be satisfied due to the power limitation of practical systems.
Different from the result  \cite{Cai2006Robust} that treats  the uncertain dynamics as a bounded total disturbance, 
the requirement for the  increment of the nonlinear dynamics in Assumption \ref{ass_sys} poses no restriction on the boundedness of uncertain dynamics. 
It is a mild condition for the systems with nonlinear uncertain dynamics.


\begin{assum}\label{ass_observable}
	There exists a positive integer $\bar N$ and a constant $\alpha>0$ such that for any $k\geq 0$, there is
	\begin{equation}\label{Observability_matrix}
	\sum_{i=1}^{N}\left[\sum_{j=k}^{k+\bar N}\Phi^T_{j,k} H_{j,i}^TR_{j,i}^{-1} H_{j,i}\Phi_{j,k}\right]\geq \alpha I_n,
	\end{equation}	
where 
\begin{align*}
\Phi_{k,k}=I_{n},\Phi_{k+1,k}=A_{k},\Phi_{j,k}=\Phi_{j,j-1}\cdots \Phi_{k+1,k}(j>k).
\end{align*}	
\end{assum}
Assumption \ref{ass_observable} is a general collective observability condition \cite{He2017Consistent}, which is a desired condition to guarantee the stability of distributed estimation algorithms. If the system is time-invariant, then Assumption \ref{ass_observable} degenerates to $(A,H)$ being observable \cite{Battistelli2014Kullback,He2017On}. Besides, if the local observability condition of an individual sensor is satisfied \cite{Cat2010Diffusion}, then Assumption \ref{ass_observable}  holds, but not vice versa.

In this paper, the topologies of the networks are assumed to be time-varying digraphs $\{\mathcal{G}_{\sigma_{k}},k\in\mathbb{N}^+\}$. 
$\sigma_{k}$ is the graph switching signal defined $\sigma_{k}:\mathbb{N}^+\rightarrow \Omega$, where $\Omega$ is the set of the underlying network topology numbers. For convenience, the weighted adjacent matrix of the digraph $\mathcal{G}_{\sigma_{k}}$ is denoted as $\mathcal{A}_{\sigma_{k}}=[a_{i,j}(k)]\in \mathbb{R}^{N\times N}$.
To analyze the time-varying topologies, we consider the infinity interval sequence of bounded, non-overlapping and contiguous time intervals $[k_{l},k_{l+1}),l=0,1,\cdots,$ with $k_{0}=0$ and $0\leq k_{l+1}-k_{l}\leq k^0$ for some integer $k^0$. On the time-varying topologies of the networks, the following assumption is in need.

\begin{assum}\label{ass_primitive}
	The digraph set $\{\mathcal{G}_{\sigma_{k}},k\in[k_{l},k_{l+1})\}$ is jointly strongly connected across the time interval $[k_{l},k_{l+1})$ and $a_{i,j}(k)\in\bar\varPsi$, $k\in\mathbb{N}^+$, where $\bar\varPsi$ is a finite set of arbitrary nonnegative numbers.
\end{assum}

Assumption \ref{ass_primitive} is on the conditions of the network topologies. Since the jointly connectedness of the time-varying digraphs admits the network is unconnected at each moment, it is quite general for the networks facing with the link failures. If the network remains connected at each moment or fixed \cite{Battistelli2014Kullback,He2017Consistent}, then Assumption \ref{ass_primitive} holds.



The objective of this paper is to estimate the extended state $X_{k}$ consisting of the original system state $x_k$ and the nonlinear dynamics $F_k$. To achieve the objective, a distributed filter is aimed to be designed for each sensor based on the information of the sensor and its neighbors.


\section{Main results}
In this section, we will propose a distributed filtering structure and study the design methods for the structure, so as to raise the extended state distributed Kalman filter of this paper. Besides, we will study the performance of this filter in terms of the boundedness of estimation covariance.

In this paper, we consider the following distributed filter structure for sensor $i$, $\forall i\in \mathcal{V}$,
\begin{equation}\label{filter_stru}
\begin{cases}
\bar X_{k,i}=A_{k-1}\hat X_{k-1,i}+D\hat u_{k-1},\\
\tilde X_{k,i}=\bar X_{k,i}+K_{k,i}(y_{k,i}- H_{k,i}\bar X_{k,i}),\\
\hat X_{k,i}=\sum_{j\in \mathcal{N}_{i}}W_{k,i,j}\tilde X_{k,j},
\end{cases}
\end{equation}
where $\bar X_{k,i}$,  $\tilde X_{k,i}$ and $\hat X_{k,i}$ are the extended state prediction,  update and  estimate of  sensor $i$ at the $k$th moment, respectively.
$K_{k,i}$ is the filtering gain matrix and $W_{k,i,j},j\in \mathcal{N}_{i}$, are the local fusion matrices. Additionally,
\begin{equation}\label{eq_u}
\begin{split}
&\hat u_{k-1}(j)=sat\big(\hat {\bar u}_{k-1}(j),\sqrt{ q_{k-1}(j)}\big),j=1,2,\cdots,p\\
&\hat {\bar u}_{k-1}=\bar F(\bar A_{k-1}\hat x_{k-1}+\bar G_{k-1}\hat F_{k-1},k)-\bar F(\hat x_{k-1},k-1)
\end{split}
\end{equation}
where $sat(\cdot)$ is the saturation function defined $sat(f,b)=\max\{\min{f,b},-b\},b>0$. It is noted that the saturation function is utilized to guarantee the boundedness of $\hat u_{k-1}$.

In the Kalman filter, the matrix $P_{k}$ stands for the estimation covariance, which can be recursively calculated. For the distributed Kalman filters, the estimation covariances are usually unaccessible, thus the following definition on consistency is introduced.
\begin{defn}(\cite{Julier1997A})
	Suppose $x_{k}$ is a random vector. Let $\hat x_{k}$ and $P_{k}$ be the estimate of $x_{k}$ and the estimate of the corresponding error covariance matrix. Then the pair ($\hat x_{k},P_{k}$) is said to be consistent (or of consistency) at the $k$th moment if
	\begin{equation*}
	E\{(\hat x_{k}-x_{k})(\hat x_{k}- x_{k})^T\}\leq P_{k}.
	\end{equation*}
\end{defn}

Regarding the filtering structure (\ref{filter_stru}), the condition for the initial estimation of each sensor is given in Assumption \ref{ass_ini}. 
\begin{assum}\label{ass_ini}
	The initial estimation of each sensor is consistent, i.e.,  
	\begin{align}
	E\{(X_{0}-\hat X_{0,i})(X_{0}-\hat X_{0,i})^T\}\leq P_{0,i},
	\end{align}	
where $P_{0,i}>0, i\in\mathcal{V}$.
\end{assum}
It is noted that Assumption \ref{ass_ini} is quite general and easy to be met, since a sufficient large $P_{0,i}$ can always be set.
Based on the filtering structure (\ref{filter_stru}), Theorem \ref{thm_W} provides a design method of fusion matrices, which can lead to the consistent estimation of each sensor.

\begin{thm}\label{thm_W}
	Under Assumptions \ref{ass_sys} and \ref{ass_ini}, for $i\in\mathcal{V},\forall \theta>0$, it follows that
\begin{equation}\label{thm_consis}
\begin{cases}
E\{(\bar X_{k,i}- X_{k})(\bar X_{k,i}- X_{k})^T\}\leq \bar P_{k,i},\\
E\{(\tilde X_{k,i}-X_{k})(\tilde X_{k,i}-X_{k})^T\}\leq \tilde P_{k,i},\\
E\{(\hat X_{k,i}-X_{k})(\hat X_{k,i}-X_{k})^T\}\leq P_{k,i},
\end{cases}
\end{equation}	
if $W_{k,i,j}$ in the filter structure (\ref{filter_stru}) are designed with
	\begin{equation}\label{eq_design_para}
	\begin{split}
	W_{k,i,j}=a_{i,j}(k)P_{k,i}^{-1} \tilde P_{k,j}^{-1},
	\end{split}
	\end{equation}
	then the pairs ($\bar X_{k,i},\bar P_{k,i}$),($\tilde X_{k,i},\tilde P_{k,i}$) and ($\hat X_{k,i},P_{k,i}$) are all consistent,
where $\bar P_{k,i},\tilde P_{k,i}$ and $P_{k,i}$ evolve according to the following recursive forms:
	         	\begin{equation}\label{eq_ite}
	         	\begin{split}
	         	&\bar P_{k,i}=(1+\theta) A_{k-1}P_{k-1,i} A_{k-1}^T+ \frac{1+\theta}{\theta} \bar Q_{k-1}+Q_{k-1},\\
	         	&\bar Q_{k-1}=4p D\cdot diag\{q_{k-1}(1),\dots,q_{k-1}(p)\}D^T,\\
	         	&\tilde P_{k,i}=(I-K_{k,i}H_{k,i})\bar P_{k,i}(I-K_{k,i}H_{k,i})^T+K_{k,i}R_{k,i}K_{k,i}^T,	\\
	         	&P_{k,i}=\bigg(\sum_{j\in \mathcal{N}_{i}}a_{i,j}(k)  \tilde P_{k,j}^{-1}\bigg)^{-1}.
	         	\end{split}
	         	\end{equation}
\end{thm}

For the filtering gain matrix $K_{k,i}$, its design  can be casted into an optimization problem given in the following lemma.
\begin{lem}\label{thm_K}
The solution of the optimization problem
\begin{align}
\min_{K_{k,i}} \tilde P_{k,i}
\end{align}
in the sense of positive definiteness is
	\begin{equation}\label{eq_design_para2}
	\begin{split}
	&K_{k,i}=\bar P_{k,i} H_{k,i}^T( H_{k,i}\bar P_{k,i}H_{k,i}^T+R_{k,i})^{-1}.
	\end{split}
	\end{equation}
\end{lem}

In Theorem \ref{thm_W}, it is shown that the  upper bounds of estimation error covariances at three typical steps can be iteratively obtained by each sensor. The bounds can not only contribute to the design of fusion weights and filtering gain, but also be used to evaluate the estimation accuracy in real time.

Summing up the results of Theorem \ref{thm_W} and Lemma \ref{thm_K}, the extended state distributed Kalman filter (ESDKF) is provided in Table \ref{ODKF2}. In the next, we will study the boundedness of the estimation error covariance of each sensor under given conditions.

\begin{table}
	\caption{Extended State Distributed Kalman Filter (ESDKF):}
	\label{ODKF2}
	\begin{tabular}{l}  
		\hline\hline
		\textbf{Prediction:}\\
		$\bar X_{k,i}=A_{k-1}\hat X_{k-1,i}+D\hat u_{k-1},$\\  
		$\bar P_{k,i}=(1+\theta) A_{k-1}P_{k-1,i} A_{k-1}^T+ \frac{1+\theta}{\theta} \bar Q_{k-1}+Q_{k-1},\forall \theta>0,$\\         
where $\hat u_{k-1}$ and $\bar Q_{k-1}$ are given in (\ref{eq_u}) and (\ref{eq_ite}), respectively.\\
		\textbf{Measurement Update:}\\
		$\tilde X_{k,i}=\bar X_{k,i}+K_{k,i}(y_{k,i}-H_{k,i}\bar X_{k,i})$\\        
		$K_{k,i}=\bar P_{k,i}H_{k,i}^T(H_{k,i}\bar P_{k,i}H_{k,i}^T+R_{k,i})^{-1}$\\
		$\tilde P_{k,i}=(I-K_{k,i}H_{k,i})\bar P_{k,i}$,\\
		\textbf{Local Fusion:} Receiving ($\tilde X_{k,j}$, $\tilde P_{k,j}$ ) from neighbors $j\in \mathcal{N}_{i}$\\
		$\hat X_{k,i}=P_{k,i}\sum_{j\in \mathcal{N}_{i}}a_{i,j}(k)\tilde P_{k,j}^{-1}\tilde X_{k,j}$,\\	
		$P_{k,i}=\bigg(\sum_{j\in \mathcal{N}_{i}}a_{i,j}(k)  \tilde P_{k,j}^{-1}\bigg)^{-1}$。	\\
		\hline
	\end{tabular}
\end{table}

\begin{thm}\label{thm_stability}
	Under Assumptions \ref{ass_sys}-\ref{ass_ini} and the condition that $\{A_{k},k\in\mathbb{N}^+\}$ belongs to a nonsingular compact set,  then there exists a positive definite matrix $\hat P$, such that
	\begin{equation}\label{thm_compare}
	E\{(\hat X_{k,i}-X_{k})(\hat X_{k,i}-X_{k})^T\}\leq P_{k,i}\leq \hat P, \forall i\in \mathcal{V}.
	\end{equation}
\end{thm}

Through Theorem \ref{thm_stability},  it can be seen that, under mild conditions including collective observability of  system and jointly strong connectedness of network topologies, the proposed filter can effectively  estimate  the extended state, consisting of the original state and the nonlinear dynamics. 

\section{Numerical Simulation}
\begin{figure}[htp]
	\centering
	\subfigure[Topology 1]{
	\begin{minipage}{3cm}
		\begin{tikzpicture}[scale=0.6, transform shape,line width=2pt]
		\tikzstyle{every node} = [circle,shade, fill=gray!30]
		\node (a) at (0, 0) {sensor 1};
		\node (b) at +(0: 1.5*2) {sensor 2};
		\node (c) at +(45: 2.1213*2) {sensor 3};
		\node (d) at +(90: 1.5*2) {sensor 4};
		\foreach \from/\to in {a/b, b/c, c/d}
		\draw [blue!30,->] (\from) -- (\to) ;
		\end{tikzpicture}
	\end{minipage}}
	\subfigure[Topology 2]{
		\begin{minipage}{3cm}
			\begin{tikzpicture}[scale=0.6, transform shape,line width=2pt]
			\tikzstyle{every node} = [circle,shade, fill=gray!30]
			\node (a) at (0, 0) {sensor 1};
			\node (b) at +(0: 1.5*2) {sensor 2};
			\node (c) at +(45: 2.1213*2) {sensor 3};
			\node (d) at +(90: 1.5*2) {sensor 4};
			\foreach \from/\to in {b/a, b/c, b/d, d/c}
			\draw [blue!30,->] (\from) -- (\to) ;
			\end{tikzpicture}
		\end{minipage}}
		\subfigure[Topology 3]{
		\begin{minipage}{3cm}
		\begin{tikzpicture}[scale=0.6, transform shape,line width=2pt]
		\tikzstyle{every node} = [circle,shade, fill=gray!30]
		\node (a) at (0, 0) {sensor 1};
		\node (b) at +(0: 1.5*2) {sensor 2};
		\node (c) at +(45: 2.1213*2) {sensor 3};
		\node (d) at +(90: 1.5*2) {sensor 4};
		\foreach \from/\to in {b/a, c/b, b/d, c/d,a/d}
		\draw [blue!30,->] (\from) -- (\to) ;
		\end{tikzpicture}
	\end{minipage}}
	\caption{The topologies of sensor networks}\label{topology}
\end{figure}
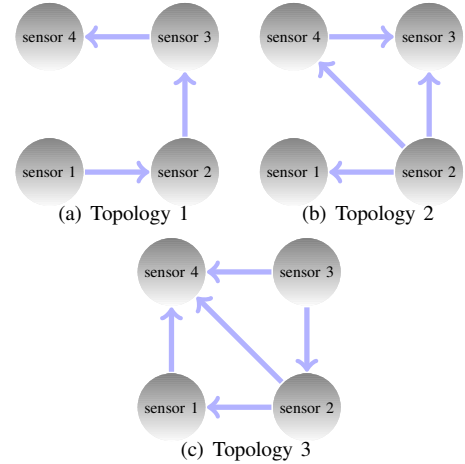

In this section, we will carry out a numerical simulation to show the effectiveness of the proposed filter.
Consider the following fourth-order time-varying system with four sensors
\begin{equation*}
	\left\{
	\begin{aligned}
		x_{k+1}&=\begin{pmatrix}
			1&0&0.1&0\\
			0&1&0&0.1\\
			0&0&1&0\\
			0&0&0&1
		\end{pmatrix}x_k+
		\frac13\begin{pmatrix}
			0\\0\\\sin\left(x_k\left(3\right)\right)+k\\\sin\left(x_k\left(4\right)\right)+k
		\end{pmatrix}\\
		&\qquad+\omega_k\\
		y_{k,i}&=\bar{H}_{k,i}x_k+v_{k,i},i=1,2,3,4
	\end{aligned}\right .,
\end{equation*}
where $x_k\in\mathrm{R}^4$ is the process and the observation matrices are
\begin{equation*}
	\left\{
	\begin{aligned}
		\bar H_{k,1}&=\begin{pmatrix}
			1&0&0&0
		\end{pmatrix}\\
		\bar H_{k,2}&=\begin{pmatrix}
			0&1&0&0
		\end{pmatrix}\\
		\bar H_{k,3}&=\begin{pmatrix}
			1&1&0&0
		\end{pmatrix}\\
		\bar H_{k,4}&=\begin{pmatrix}
			0&0&0&0
		\end{pmatrix}\\
	\end{aligned}\right ..
\end{equation*}
Additionally, the network communication topologies, assumed as directed and time-varying, are illustrated in Fig. \ref{topology} with adjacent matrix selected from
\begin{equation*}
	\left\{\begin{aligned}
		\mathcal{A}_1&=\begin{pmatrix}
			1&0&0&0\\
			0.5&0.5&0&0\\
			0&0.5&0.5&0\\
			0&0&0.5&0.5
		\end{pmatrix}\\
		\mathcal{A}_2&=\begin{pmatrix}
			0.5&0.5&0&0\\
			0&1&0&0\\
			0&0.3&0.4&0.3\\
			0&0.5&0&0.5
		\end{pmatrix}\\
		\mathcal{A}_3&=\begin{pmatrix}
			0.5&0.5&0&0\\
			0&0.5&0.5&0\\
			0&0&1&0\\
			0.25&0.25&0.25&0.25
		\end{pmatrix}\\
	\end{aligned}\right ..
\end{equation*}
And the gragh switching signal $\sigma_k$ is
\begin{equation*}
	\left\{
	\begin{aligned}
		\sigma_k&=1,k=1:5,16:20,\cdots\\
		\sigma_k&=2,k=6:10,21:25,\cdots\\
		\sigma_k&=3,k=11:15,26:30,\cdots
	\end{aligned}\right ..
\end{equation*}

Here, it is assumed that the process noise covariance matrix $Q=diag\{4,4,1,1\}$, and the whole measurement noise covariance matrix $R=diag\{90,90,90,90\}$, and the parameter $\theta=0.1$.
The initial value has zero mean and covariance matrix being $100I_4$, and the initial estimation setting is $\hat X_{i,0}=0$ and $P_{i,0} = 100I_4, \forall i=1,2,3,4$.
Next, we conduct the numerical simulation through Monte Carlo experiment, in which 500 runs for the proposed algorithm are implemented. The average performance function is defined $MSE_k=\frac14\sum_{i=1}^4 {MSE_{k,i}}$, where
\begin{equation*}
	MSE_{k,i}=\frac1{500}\sum_{j=1}^{500}{\left(X_{k,i}^j-\hat{X}_{k,i}^j\right)^T\left(X_{k,i}^j-\hat{X}_{k,i}^j\right)},
\end{equation*}
and $\hat{X}_{k,i}^j$ is the extended state estimate of the $j$th run of sensor $i$ at the $k$th moment.
Besides, we define $P_{k}=\frac14\sum_{i=1}^4 {P_{k,i}}$.

\begin{figure}
	\centering
	\includegraphics[scale=0.6]{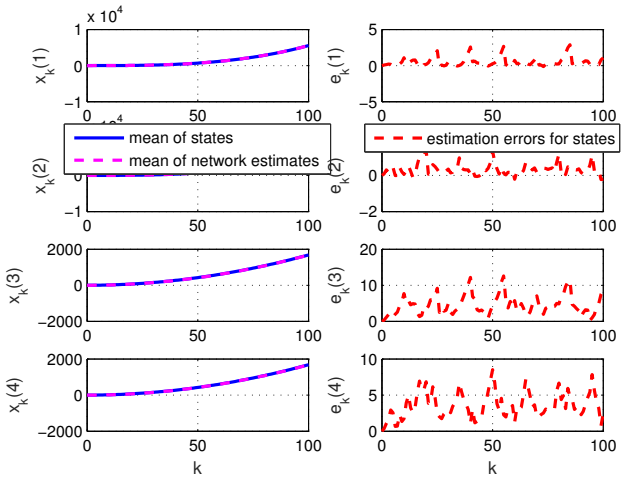}
	\caption {Tracking performance of ESDKF}
	\label{fig:trace}
\end{figure}

\begin{figure}
	\centering
	\includegraphics[scale=0.6]{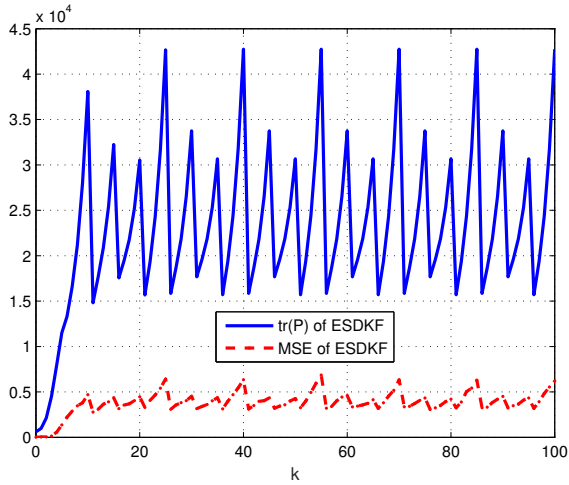}
	\caption {Performance evaluation of ESDKF}
	\label{fig:MSE}
\end{figure}

The simulation results based on above parameter setting are gived in Fig. \ref{fig:trace} and Fig. \ref{fig:MSE}.
Fig. \ref{fig:trace} shows the tracking performance for the system states and the estimation errors, from which one can see the algorithm can effectively estimate the overall state in the sense of mean value. The comparison between $MSE_{k,i}$ and $tr(P_{k,i})$ is depicted in Fig. \ref{fig:MSE}, where it  can be seen that the estimation errors of the proposed
algorithm keep stable in the given period and the consistency of each sensor remains.

\section{Conclusion}\label{Conclusion}
In this paper, the  distributed state estimation problem was studied for a  class of discrete-time stochastic systems with nonlinear uncertain dynamics over time-varying topologies.  Through extending the  original state and the nonlinear dynamics to  a new state vector, an extended system was constructed.
Then, a design method for the filtering gain and fusion matrices was proposed. It was shown that the proposed filter was consistent, i.e., the upper bound of estimation covariance can be provided in real time. 
It was proven that the  estimation covariance of the filter was bounded under rather mild assumptions, i.e., collective observability of the system and jointly strong connectedness of network topologies.
 Through numerical simulations, the effectiveness of the distributed filter was testified.


\section*{Acknowledgment}
The authors would like to thank the financial support in part by the NSFC61603380,  the National Key Research and Development Program of China (2016YFB0901902), the National Basic Research Program of China under Grant No. 2014CB845301.



%

%
%

\bibliography{references_filtering}

\begin{thebibliography}{10}

\bibitem{Pfaff2017Information}
F.~Pfaff, B.~Noack, U.~D. Hanebeck, F.~Govaers, and W.~Koch, ``Information form
  distributed \protect{Kalman} filtering with explicit inputs,'' in {\em
  International Conference on Information Fusion}, pp.~1--8, 2017.

\bibitem{Chong2014Comparison}
C.~Y. Chong, S.~Mori, F.~Govaers, and W.~Koch, ``Comparison of tracklet fusion
  and distributed \protect{Kalman} filter for track fusion,'' in {\em
  International Conference on Information Fusion}, 2014.

\bibitem{Khan2014Collaborative}
U.~A. Khan and A.~Jadbabaie, ``Collaborative scalar-gain estimators for
  potentially unstable social dynamics with limited communication,'' {\em
  Automatica}, vol.~50, no.~7, pp.~1909--1914, 2014.

\bibitem{Das2015Distributed}
S.~Das and J.~M.~F. Moura, ``Distributed \protect{Kalman} filtering with
  dynamic observations consensus,'' {\em IEEE Transactions on Signal
  Processing}, vol.~63, no.~17, pp.~4458--4473, 2015.

\bibitem{Cat2010Diffusion}
F.~S. Cattivelli and A.~H. Sayed, ``Diffusion strategies for distributed
  \protect{Kalman} filtering and smoothing,'' {\em IEEE Transactions on
  Automatic Control}, vol.~55, no.~9, pp.~2069--2084, 2010.

\bibitem{Mahmoud2013Distributed}
M.~S. Mahmoud and H.~M. Khalid, ``Distributed \protect{Kalman} filtering: a
  bibliographic review,'' {\em IET Control Theory \& Applications}, vol.~7,
  no.~4, pp.~483--501, 2013.

\bibitem{He2017Consistent}
X.~He, W.~Xue, and H.~Fang, ``Consistent distributed state estimation with
  global observability over sensor network,'' {\em Automatica}, vol.~92,
  pp.~162--172, 2018.

\bibitem{Battistelli2015Consensus}
G.~Battistelli, L.~Chisci, G.~Mugnai, A.~Farina, and A.~Graziano,
  ``Consensus-based linear and nonlinear filtering,'' {\em IEEE Transactions on
  Automatic Control}, vol.~60, no.~5, pp.~1410--1415, 2015.

\bibitem{yang2017stochastic}
W.~Yang, C.~Yang, H.~Shi, L.~Shi, and G.~Chen, ``Stochastic link activation for
  distributed filtering under sensor power constraint,'' {\em Automatica},
  vol.~75, pp.~109--118, 2017.

\bibitem{Reif1999Stochastic}
K.~Reif, S.~Gunther, E.~Yaz, and R.~Unbehauen, ``Stochastic stability of the
  discrete-time extended \protect{Kalman} filter,'' {\em IEEE Transactions on
  Automatic Control}, vol.~44, no.~4, pp.~714--728, 1999.

\bibitem{Julier2004Unscented}
S.~J. Julier and J.~K. Uhlmann, ``Unscented filtering and nonlinear
  estimation,'' {\em Proceedings of the IEEE}, vol.~92, no.~3, pp.~401--422,
  2004.

\bibitem{yang2008non}
G.~Yang and W.~Che, ``Non-fragile $\protect{H}_{\infty}$ filter design for
  linear continuous-time systems,'' {\em Automatica}, vol.~44, no.~11,
  pp.~2849--2856, 2008.

\bibitem{ding2012distributed}
D.~Ding, Z.~Wang, H.~Dong, and H.~Shu, ``Distributed $\protect{H}_{\infty}$
  state estimation with stochastic parameters and nonlinearities through sensor
  networks: the finite-horizon case,'' {\em Automatica}, vol.~48, no.~8,
  pp.~1575--1585, 2012.

\bibitem{calafiore2005reliable}
G.~Calafiore, ``Reliable localization using set-valued nonlinear filters,''
  {\em IEEE Transactions on systems, man, and cybernetics-part A: systems and
  humans}, vol.~35, no.~2, pp.~189--197, 2005.

\bibitem{Bai2017reliable}
W.~Bai, W.~Xue, Y.~Huang, and H.~Fang, ``On extended state based
  \protect{Kalman} filter design for a class of nonlinear time-varying
  uncertain systems,'' {\em Science China Information Sciences}, vol.~61,
  no.~4, p.~042201, 2018.

\bibitem{Zhang2018}
X.~Zhang, W.~Xue, H.~Fang, and X.~He, ``On extended state based
  \protect{Kalman-Bucy} filter,'' in {\em IEEE 7th Data Driven Control and
  Learning Systems Conference}, 2018.

\bibitem{He2018Dis}
X.~He, X.~Zhang, W.~Xue, and H.~Fang, ``Distributed \protect{Kalman} filter for
  a class of nonlinear uncertain systems: An extended state method,'' {\em DOI:
  10.13140/RG.2.2.34429.26088}, 2018.

\bibitem{Cai2006Robust}
Z.~Cai, M.~S.~D. Queiroz, and D.~M. Dawson, ``Robust adaptive asymptotic
  tracking of nonlinear systems with additive disturbance,'' {\em IEEE
  Transactions on Automatic Control}, vol.~51, no.~3, pp.~524--529, 2006.

\bibitem{Battistelli2014Kullback}
G.~Battistelli and L.~Chisci, ``Kullback-\protect{Leibler} average, consensus
  on probability densities, and distributed state estimation with guaranteed
  stability,'' {\em Automatica}, vol.~50, no.~3, pp.~707--718, 2014.

\bibitem{He2017On}
X.~He, C.~Hu, W.~Xue, and H.~Fang, ``On event-based distributed
  \protect{Kalman} filter with information matrix triggers,'' in {\em IFAC
  World Congress}, pp.~14873--14878, 2017.

\bibitem{Julier1997A}
S.~J. Julier and J.~K. Uhlmann, ``A non-divergent estimation algorithm in the
  presence of unknown correlations,'' in {\em Proceedings of the American
  Control Conference}, pp.~2369--2373, 1997.

\end{thebibliography}
\bibliographystyle{ieeetr}

\appendices

\end{document}